\documentstyle[prb,preprint,aps]{revtex}
\begin{document}
\renewcommand{\thefootnote}{\fnsymbol{footnote}}
\draft

\title{NMR relaxation rates for the spin-1/2 Heisenberg chain}

\author{Anders W. Sandvik\cite{abo}}
\address{National High Magnetic Field Laboratory, Florida State
University, 1800 E. Paul Dirac Dr., Tallahassee, Florida 32306 }

\maketitle

\narrowtext

\begin{abstract}
The spin-lattice relaxation rate $1/T_1$ and the spin echo decay rate
$1/T_{2G}$ for the spin-$1\over 2$ antiferromagnetic Heisenberg chain
are calculated using quantum Monte Carlo and maximum entropy analytic
continuation. The results are compared with recent analytical
calculations by Sachdev. If the nuclear hyperfine form factor $A_q$ is
strongly peaked around $q=\pi$ the predicted low-temperature behavior
[$1/T_1 \sim \ln{^{1/2}(1/T)}$, $1/T_{2G} \sim \ln{^{1/2}(1/T)}/\sqrt{T}$]
extends up to temperatures as high as $T/J \approx 0.5$. If $A_q$ has
significant weight for $q \approx 0$ there are large contributions from
diffusive long-wavelength processes not taken into account in the
theory, and very low temperatures are needed in order to observe the
asymptotic $T \to 0$ forms.
\end{abstract}
\vfill\eject

The one-dimensional spin-$1\over 2$ antiferromagnetic Heisenberg hamiltonian,
\begin{equation}
\hat H=J\sum\limits_{i} \vec S_i \cdot \vec S_{i+1} ,
\label{heisenberg}
\end{equation}
is relevant as
a starting point for understanding the magnetic properties of many
quasi one-dimensional materials. Examples include
CuCl$\cdot$2N(C$_5$H$_5$),\cite{endoh} KCuF$_3$,\cite{tennant} and
several tetracyanoquinodimethan (TCNQ) charge transfer
salts.\cite{epstein,ehrenfreund} NMR and NQR are commonly used
techniques for studying the spin dynamics of materials such as those
listed above. The low-frequency dynamic susceptibility is accessible
through the spin-lattice relaxation rate $1/T_1$ and the spin echo decay
rate $1/T_{2G}$. Theoretical results for the temperature dependence of
both these rates were recently obtained by Sachdev,\cite{sachdev} using a
form for the dynamic susceptibility first derived by Schulz
\cite{schulz} using the Bosonization method. Neglecting logarithmic
corrections, $1/T_1$ is predicted to be constant at low temperature,
and $1/T_{2G}$ is predicted to diverge as $T^{-1/2}$. With logarithmic
corrections taken into account both rates acquire a factor ln$^{1/2}(1/T)$.
These results are expected to be valid only for temperatures $T \ll J$,
and it is important to verify their validity as well as to obtain results
also at higher temperatures. Here results are presented for $1/T_1$ and
$1/T_{2G}$ computed using quantum Monte Carlo (QMC) simulations of chains of
up to 1024 spins. $1/T_{2G}$ is related to static susceptibilities directly
computable in the simulations. The dynamic susceptibility required for
extracting $1/T_1$ is calculated in imaginary time and continued to real
frequency using the maximum entropy method.~\cite{maxent1,maxent2}

The results for the temperature dependence of both $1/T_1$ and
$1/T_{2G}$ at low temperatures are in good agreement with Sachdev's
predictions. At higher temperatures diffusive modes not taken into
account in the theory cause significant deviations. If the nuclear
hyperfine form factor has large weight at long wavelengths very low
temperatures are needed for the asymptotic forms to apply, and they
may then be difficult to observe experimentally. The results presented
here should be useful for comparisons with experiments also at higher
temperatures.

The NMR spin-lattice relaxation rate is given by~\cite{moriya}
\begin{equation}
{1\over T_1} = \hbox{$1\over N$} \sum\limits_{\alpha} \sum\limits_q
| A^{\alpha}_q | ^2 S(q,\omega \to 0),
\end{equation}
where $A^{\alpha}_q$ is the hyperfine form factor, and
$\alpha$ denotes the two axes perpendicular to the external field
direction. $S(q,\omega)$ is the dynamic structure factor, which is
related to the imaginary part of the dynamic spin susceptibility according to
$S(q,\omega)=\chi''(q,\omega)/(1-\hbox{e}^{-\beta\omega})$, where
$\beta = 1/k_BT$. Here an isotropic form factor $A^\alpha_q = A_q$
will be assumed. Defining
\begin{equation}
S_A(\omega) = \hbox{$1\over N$} \sum\limits_q | A_q | ^2 S(q,\omega),
\end{equation}
the spin-lattice relaxation rate is then obtained as
$1/T_1 = 2S_A(\omega \to 0)$. For the numerical calculations carried out
here it is more convenient to work directly with the hyperfine coupling
$A(r)$ in coordinate space. Define
\begin{equation}
C_A(\tau) =
\sum\limits_{i,j} A(r_i)A(r_j) \langle S^z_i (\tau)S^z_j(0) \rangle ,
\end{equation}
where $S^z_i (\tau) = \hbox{e}^{\tau \hat H} S^z_i \hbox{e}^{-\tau H}$.
With $C_A(\tau)$ calculated numerically, $S_A(\omega)$ can be obtained
by inverting the relation
\begin{equation}
C_A(\tau) = {1\over \pi} \int\limits_{-\infty}^\infty d\omega S_A(\omega)
\hbox{e}^{-\tau\omega}
\label{inversion}
\end{equation}
using the maximum entropy technique.\cite{maxent1,maxent2} This method
is described in detail in Ref. \onlinecite{maxent2}, and was recently
applied in a calculation of the spin-lattice relaxation rate of the
two-dimensional Heisenberg model.\cite{rates2d}

The gaussian component of the spin echo decay rate is related to the
the nuclear spin-spin interactions mediated by the electrons.
Under conditions discussed by Pennington
and Slichter~\cite{pennington}
\begin{equation}
{1\over T_{2G}} =
\Bigl [\hbox{$1\over 2$} \sum\limits_{x \not= 0}
J_z^2 (0,x) \Bigr ] ^{1/2},
\end{equation}
where $J_z(x_1,x_2)$ is the $z$-component of the induced
interaction between nuclei at $x_1$ and $x_2$:
\begin{equation}
J_z(x_1,x_2) = -\hbox{$1\over 2$}
\sum\limits_{i,j} A(x_1 - r_i)A(x_2 - r_j)\chi (i-j) .
\end{equation}
The static susceptibility $\chi (i-j)$ is given by the Kubo formula
\begin{equation}
\chi (i-j) = \int\limits_0^\beta d\tau \langle S^z_i (\tau )
                  S^z_j (0) \rangle .
\label{statsus}
\end{equation}

The hyperfine interaction $A(r)$ is normally very short ranged. Here
a situation is considered where the nuclei studied reside at the sites of the
electronic spins modeled by the hamiltonian (\ref{heisenberg}). The
hyperfine coupling is assumed to have a direct contact term of strength
$A(0)$, and a transferred nearest-neighbor term of strength $A(1)$.
Results are presented for several values of the ratio $R=A(1)/A(0)$.

A stable inversion of the relation (\ref{inversion})
requires that $C_A(\tau)$ is known to very high accuracy. Here a quantum
Monte Carlo method based on stochastic series expansion\cite{qmc}
(a generalization of Handscomb's method\cite{handscomb}) is used.
This technique is free from systematical errors of the ``Trotter
break-up'' used in standard methods.\cite{wlmethod}
The imaginary time correlation functions needed have been calculated to within
relative statistical errors of only 10$^{-4}$ or lower for
temperatures down to $T = J/8$. This high accuracy is
required for obtaining a reliable estimate of $S_A(\omega \to 0)$.
The static susceptibilities (\ref{statsus}) are computed directly in
the QMC simulation, and hence the calculation of $1/T_{2G}$ is not hampered
by potential problems associated with analytic continuation. Accurate
results for $1/T_{2G}$ have been obtained at temperatures as low as
$T = J/32$ for systems of up to 1024 sites, which is large enough for
finite-size effects to be completely negligible.

In order to test the accuracy of a calculation of $1/T_1$ by analytic
continuation of QMC data, complete diagonalizations of 16-site chains were
also carried out. Comparisons of $S(\omega) = {1\over N}\sum_q S(q,\omega)$
obtained in these calculations with numerically continued QMC data
are shown in Fig.~1. The maximum entropy method requires a ``default
model'' which defines the zero of entropy.\cite{maxent1,maxent2} In all
calculations presented here a flat default model was used.
Exact diagonalization gives $S(\omega)$ as a finite number of delta
functions and their corresponding weights. Here the results are plotted
as histograms in order to
facilitate comparison with the maximum entropy result. The jagged structure
of the exact diagonalization result, which is due to the small size of
the system, cannot be reproduced by the maximum entropy method. The
results do however represent reasonable frequency averages. Note that
even the high-frequency behavior is obtained quite accurately. Clearly a
$16$-site system is not large enough for extracting the low-frequency
behavior at low temperatures. The $1/T_1$ results presented below are for
systems of 256 spins, and the accuracy of the imaginary time data used
for the analytic continuation is even higher than the data used for the
16-site results shown in Fig.~1. Comparisons with results obtained for
128 spins indicate that there are no significant finite size effects at
the temperatures considered.

Both fluctuations of the uniform ($q\approx 0$) and staggered
($q\approx \pi$) magnetization contribute to the NMR rates of
half-integer spin chains. At low temperatures the staggered contribution
dominates.\cite{sachdev} Neglecting the uniform fluctuations, Sachdev
obtained the asymptotic low-temperature forms
(in units where $\hbar = k_B = 1$)
\begin{mathletters}
\begin{eqnarray}
{1\over T_1} & = & A^2_\pi {\pi D \over c} ,\\
\label{ratea}
{1\over T_{2G}} & = & A^2_\pi {I D \over 4} \sqrt{1\over 2cT},
\label{rateb}
\end{eqnarray}
\label{rates}
\end{mathletters}
where $I \approx  8.4425$, and $c$ is the spinon velocity,
which for spin-$1\over2$ is $c= {\pi\over 2}$. $D$ is the prefactor of
the asymptotic equal-time spin-correlation function, which is not known
accurately.\cite{hallberg} The marginally irrelevant operator present
for the critical spin chains has not been taken into account in the
derivation of the above forms. This is expected
to lead to a multiplicative correction $\ln{^{1/2}(\Lambda /T)}$ for
both $1/T_1$ and $1/T_{2G}$.\cite{sachdev}~ Hence, the ratio
$T_{2G}/(\sqrt{T}T_1)$ should be a constant, even with logarithmic
corrections included.

If $1/T_{2G} \sim \hbox{ln}^{1/2}(\Lambda /T)/\sqrt{T}$ as predicted by
Sachdev, $T(1/T_{2G})^2$ should be a linear function of ln$(J/T)$. In
Fig.~2, $T(A^2_\pi T_{2G})^{-2}$ is graphed versus $\ln{(J/T)}$ for several
values of the hyperfine coupling ratio $R=A(1)/A(0)$. In cases where
the corresponding $A_q$ is peaked around $q = \pi$ ($R < 0$) a linear
behavior is seen in a wide temperature regime. The points for $R=-0.25$
and $R=-0.5$ nearly coincide at low $T$, indicating that the  $q\approx \pi$
contributions almost completely dominate the behavior in both cases.
A line fit to the $R=-0.5$ points gives $\Lambda = 0.92J$ and the
amplitude $D=0.080$ in (\ref{rateb}). For $R > 0$ contributions from
$q \approx 0$ rapidly become important at high temperatures, and for
large values of $R$ the asymptotic behavior can only be observed at very low
temperatures.

Results for $1/T_1$ divided by $A_\pi^2$ are shown in Fig.~3.
The expected weak (logarithmic) increase as $T$ decreases can be seen
below $T/J = 0.5$ if $R$ is large and negative, so that $q \approx \pi$
processes dominate $S_A (\omega \to 0)$ even at relatively high temperatures.
For $R > 0$, $1/T_1$ decreases with decreasing $T$ down to quite low
temperatures --- for $R=0.25$ this behavior extends down to the lowest
temperature studied. The enhancement of $1/T_1$ at high $T$ is caused by the
diffusive $q \approx 0$ processes not taken into account in the forms
(\ref{rates}).\cite{sachdev} In order to more clearly determine the
importance of these modes one can study the ratio
\begin{equation}
{S_{q<\pi/2}(\omega \to 0) \over S(\omega \to 0)} =
{\sum_{q< \pi/2} S(q,\omega \to 0) \over \sum_{q} S(q,\omega \to 0)},
\end{equation}
which is graphed versus the temperature in Fig.~4
(these calculations were carried out on systems of 128 spins).
At $T=J$ the $q < \pi/2$ contribution is approximately 50\%, and
decreases rapidly at lower temperatures.
These results confirm Sachdev's conclusion \cite{sachdev} that
the $q \approx 0$ contribution to $1/T_1$ is negligible in the
limit $T\to 0$.

Returning now to the results shown in Fig.~3, there are not enough
low-temperature data to extract the asymptotic temperature dependence of
$1/T_1$. The results are, however, consistent with a divergence of the
predicted form ln$^{1/2} (\Lambda /T)$ with the same $\Lambda=0.92J$ as
was found above for $1/T_{2G}$. The amplitude needed in Eq.
(\ref{ratea}) is then $D\approx 0.14$, which is significantly larger
than the amplitude extracted from $1/T_{2G}$ above. Hence, the ratio
$T_{2G}/(\sqrt{T}T_1)$ is different from Sachdev's prediction. The ratio
is graphed versus temperature in Fig.~5. For $R < 0$ it is indeed almost
constant below $T/J \approx 0.5$, whereas for $R \ge 0$ there is a significant
temperature dependence down to the lowest temperatures considered.
For positive $R$ there is a sharp maximum in
$T_{2G}/(\sqrt{T}T_1)$, arising from the minimum in $1/T_{2G}$ seen in
Fig.~2. The $R=-0.5$ result for $T_{2G}/(\sqrt{T}T_1)$ at low temperatures
is approximately $3.0-3.1$, which is almost a factor $2$ larger than what
is obtained from Eqs. (\ref{rates}).

In summary, the NMR rates $1/T_1$ and $1/T_{2G}$ have been calculated
for the spin-1/2 Heisenberg model, using quantum Monte Carlo and
maximum entropy analytic continuation. The temperature dependence at low
temperature is in good agreement with Sachdev's recent theoretical
results, which include only the contributions from staggered
magnetization fluctuations. At high temperature damped $q \approx 0$
modes are important, and can dominate the NMR rates if the
hyperfine form factor has large weight at long wavelengths. In such cases
very low temperatures are needed to observe the asymptotic forms.
In many real systems effects of interchain couplings may become
important before the asymptotic regime is reached, and the low-temperature
forms may therefore not be easily observed. The results here
should then be useful for determining the relevance of a description
by the one-dimensional Heisenberg model based upon measurements at higher
temperatures. It can be noted that early NMR experiments\cite{ehrenfreund}
on (NMP)(TCNQ) indicate a behavior of $1/T_1$ similar to the result shown in
Fig.~3 for a small positive hyperfine ratio $A(1)/A(0)$, with no
indication of a low-temperature increase down to $T \approx 0.1J$.

It will be interesting to apply the techniques used here to calculate
the NMR rates of other one-dimensional systems. Work on coupled spin chains
is in progress.\cite{chains} Itinerant electrons described by
one-dimensional Hubbard-type models, including electron-phonon
interactions, can also be studied.

I would like to thank E. Dagotto, S. Haas, and D. Scalapino for
useful discussions. Most of the computations were carried out on a
cluster of DEC Alpha workstations at the Supercomputer Computations
Research Institute at Florida State University. This work was supported
by the Office of Naval Research under Grant No. ONR N00014-93-0495.

\begin{figure}
FIG.~1. QMC and exact diagonalization results for
$S(\omega)={1\over N}\sum_q S(q,\omega)$ of a 16-site chain. The exact
results are plotted as histograms, and the curves are numerically
continued QMC imaginary time correlation functions.
\end{figure}

\begin{figure}
FIG.~2. Results for the spin echo decay rate, graphed as
$T(A_\pi^2 T_{2G})^{-2}$ vs. ln$(J/T)$ for different hyperfine
ratios $R=A(1)/A(0)$.
\end{figure}

\begin{figure}
FIG.~3. Spin-lattice relaxation rates vs. temperature for different
hyperfine ratios $R=A(1)/A(0)$. The solid curve is of the form
ln$(\Lambda /T)$, with $\Lambda = 0.92J$.
\end{figure}

\begin{figure}
FIG.~4. The long-wavelength contribution to
$S(\omega \to 0)=\sum_q S(q,\omega \to 0)$ vs. temperature.
\end{figure}

\begin{figure}
FIG.~5. The ratio $T_{2G}/\sqrt{T}T_1$ vs. $T$ for
different hyperfine ratios $R=A(1)/A(0)$.
\end{figure}

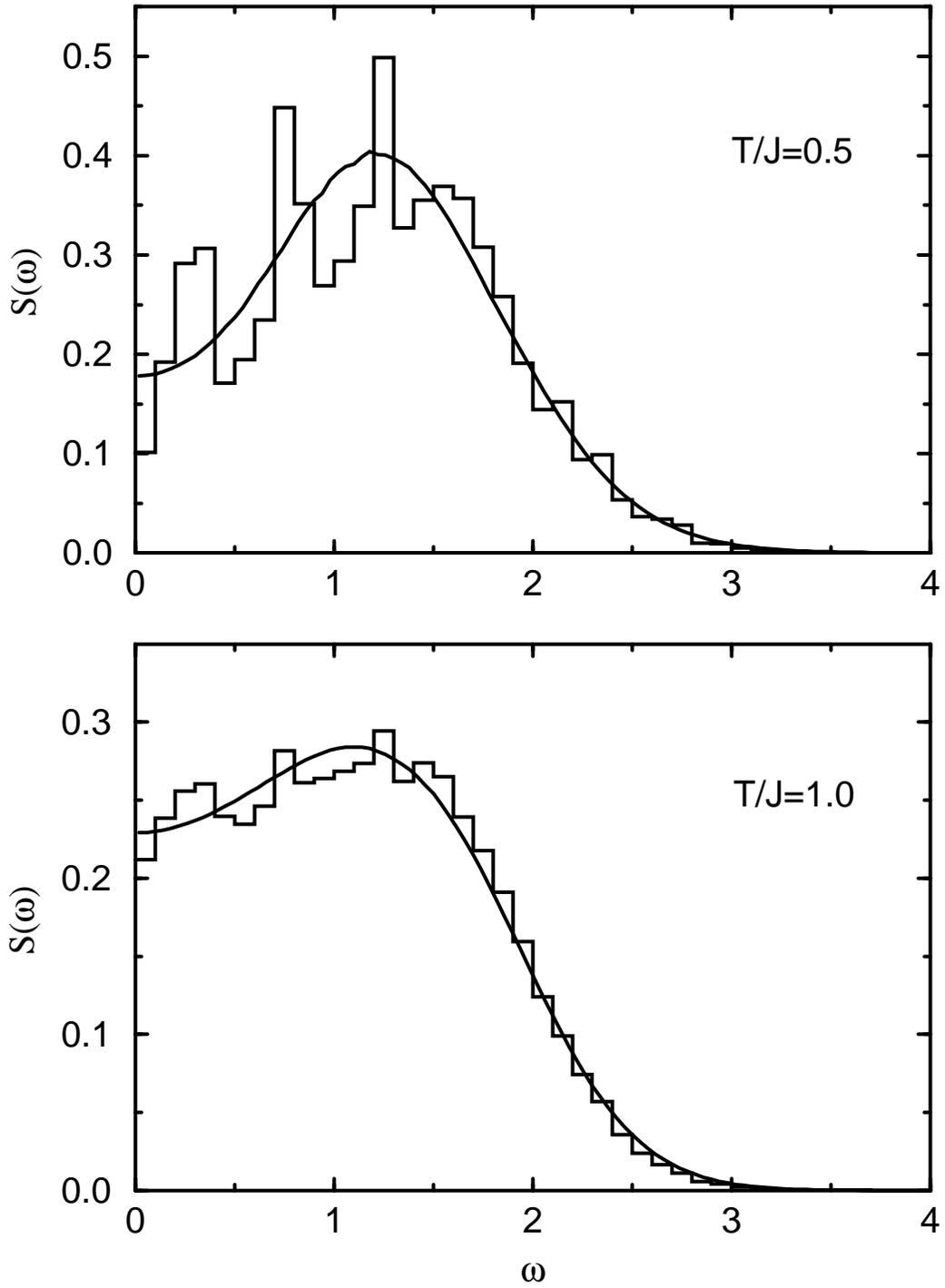
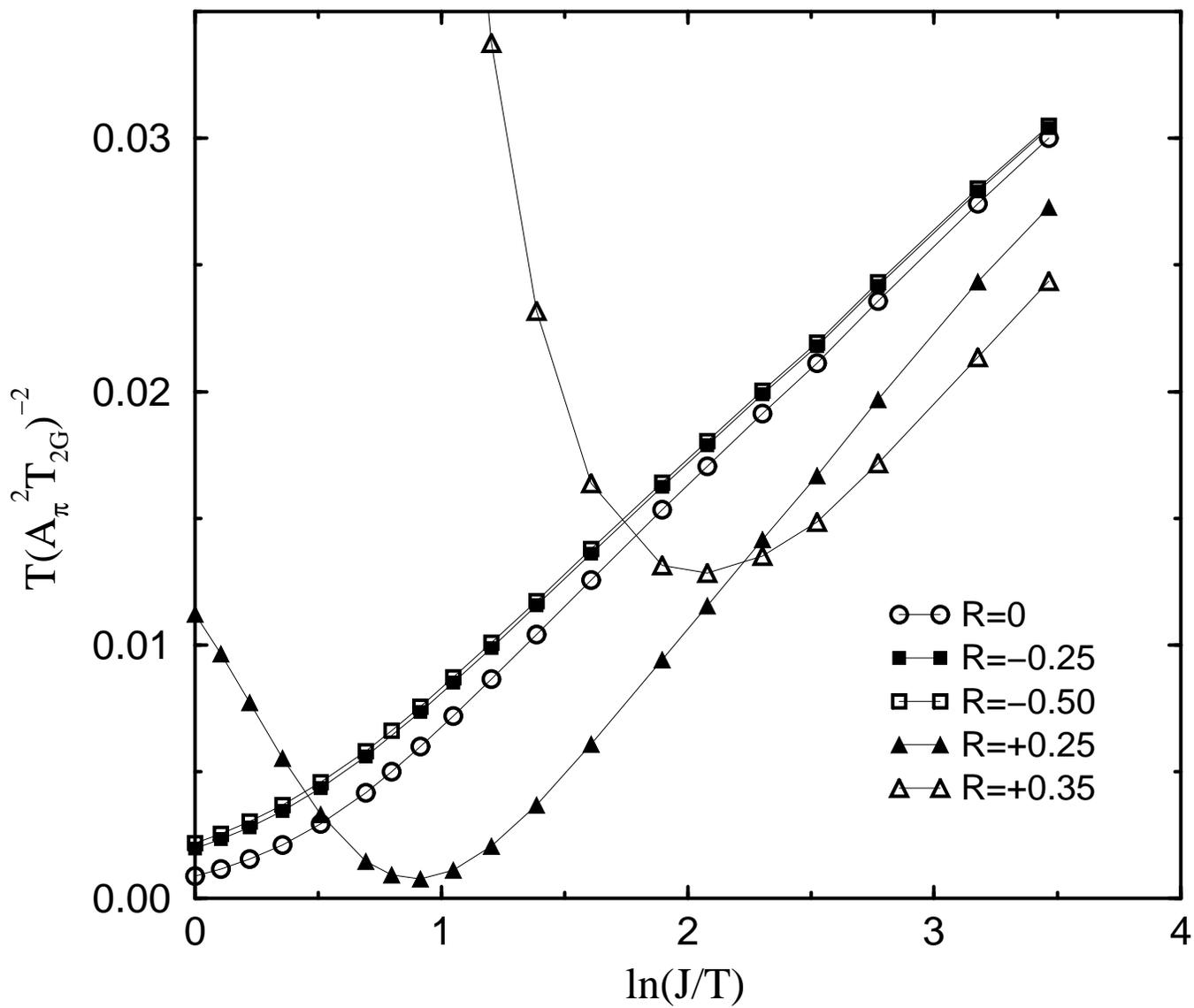
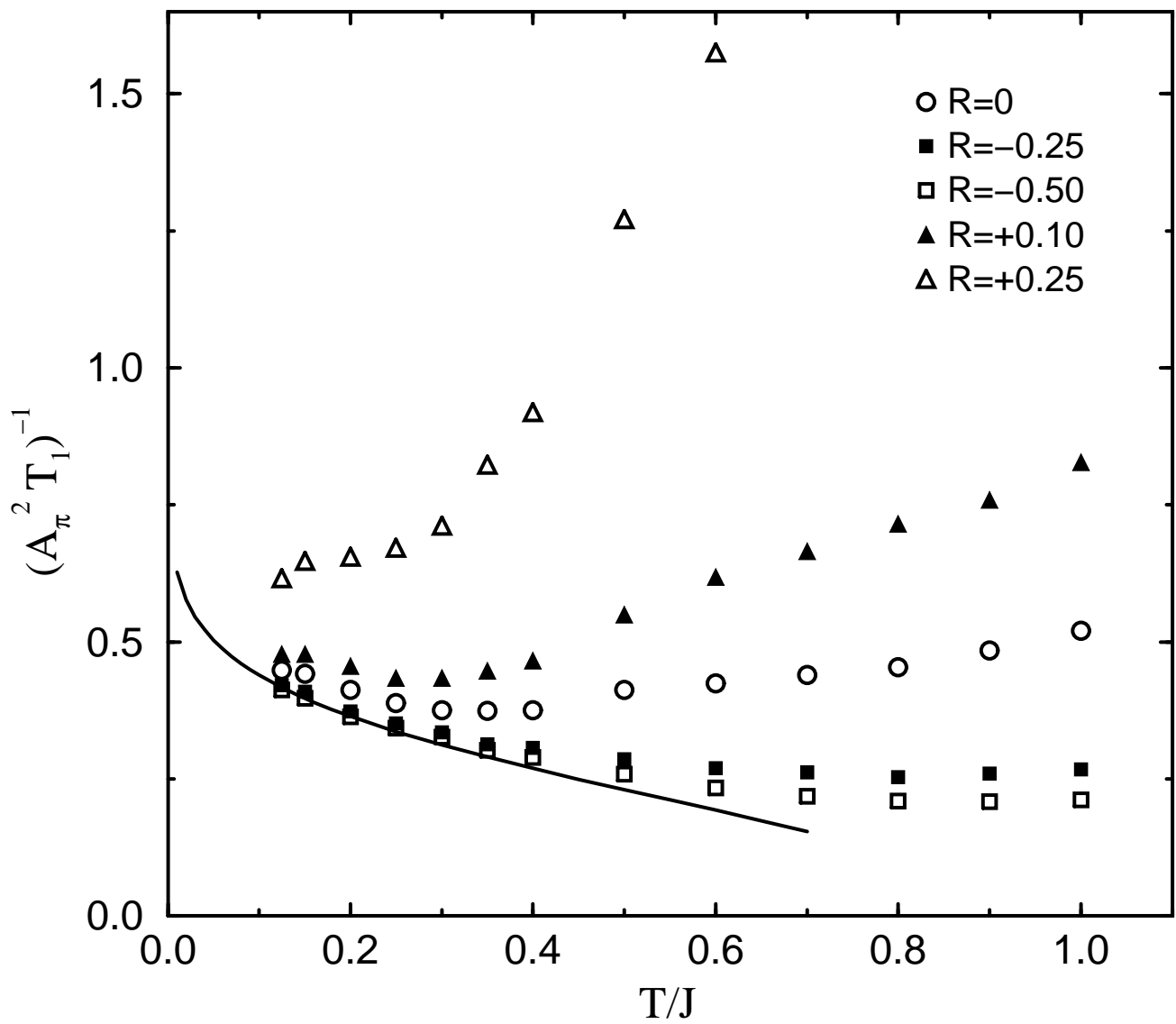
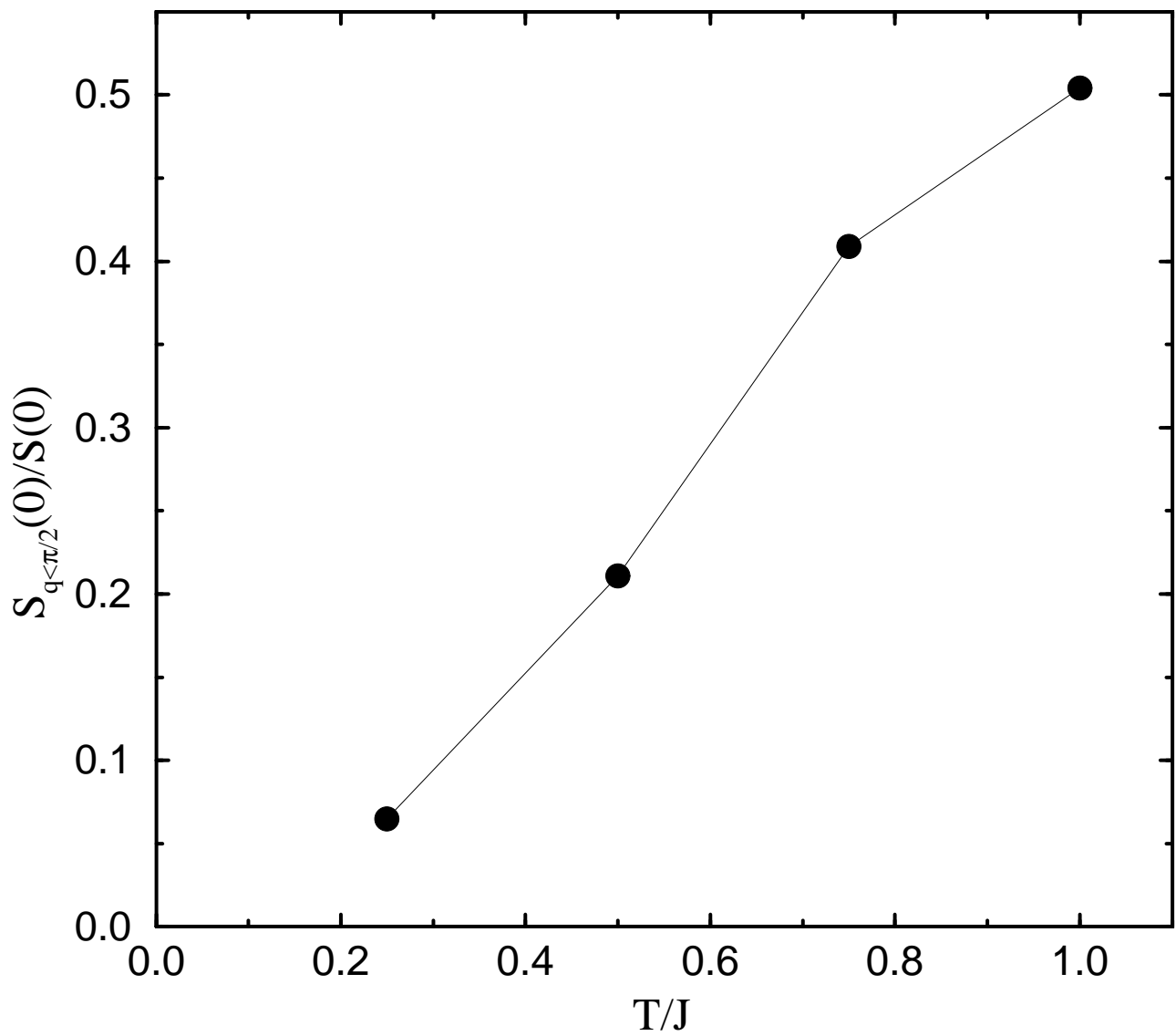
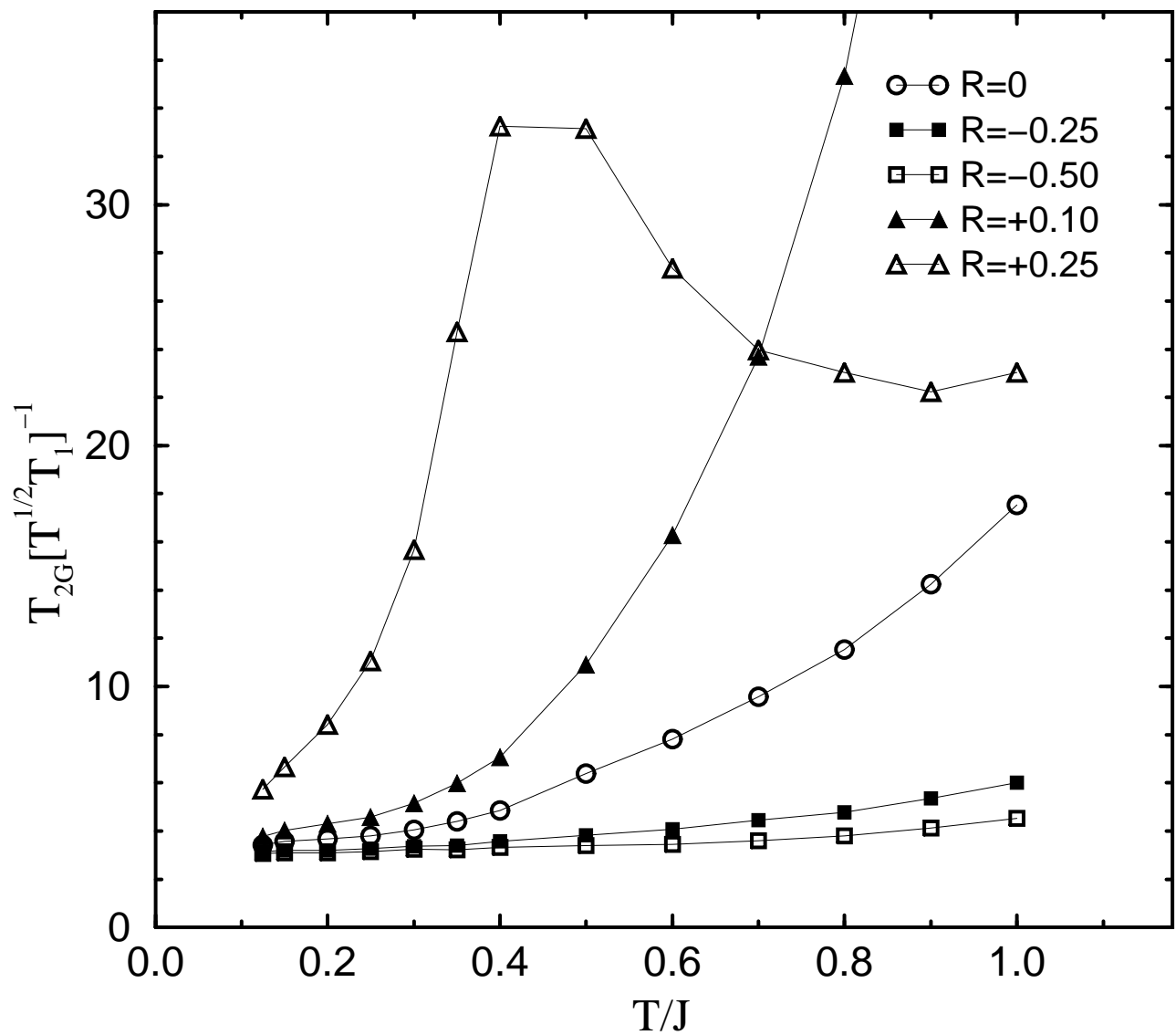
\end{document}